\documentclass[aps,prl,twocolumn,groupedaddress]{revtex4}

\usepackage{graphicx}

\begin{document}

%\preprint{}

\title{Quantum dynamics in a camel-back potential of a dc SQUID}


\author{E. Hoskinson$^1$, F. Lecocq$^1$, N. Didier$^2$, A. Fay$^1$,
  F. W. J. Hekking$^2$, W. Guichard$^1$ and O. Buisson$^1$}

\affiliation{$^1$Institut N\'eel, C.N.R.S.- Universit\'e
Joseph Fourier, BP 166, 38042 Grenoble-cedex 9, France}

\affiliation{$^2$LPMMC, C.N.R.S.- Universit\'e
Joseph Fourier, BP 166, 38042 Grenoble-cedex 9, France}

\author{R. Dolata, B. Mackrodt and A. B. Zorin}
\affiliation{Physikalisch-Technische Bundesanstalt, Bundesallee 100,
  38116 Braunschweig, Germany}

%\thanks{}
\affiliation{}


\date{\today}

\begin{abstract}
  We investigate the quantum dynamics of a quadratic-quartic
  anharmonic oscillator formed by a potential well between two
  potential barriers. We realize this novel potential shape with a
  superconducting circuit comprised of a loop interrupted by two
  Josephson junctions, with near-zero current bias and flux bias near
  half a flux quantum. We investigate escape out of the central well,
  which can occur via tunneling through either of the two barriers,
  and find good agreement with a generalized double-path macroscopic
  quantum tunneling theory. We also demonstrate that this system
  exhibits an ``optimal line'' in current and flux bias space along
  which the oscillator, which can be operated as a phase qubit, is
  insensitive to decoherence due to low-frequency current
  fluctuations.
\end{abstract}

\pacs{85.25.Cp, 85.25.Dq, 03.67.Lx}
%\keywords{}

\maketitle

\begin{figure}[htb*]
\includegraphics[width=3.3in]{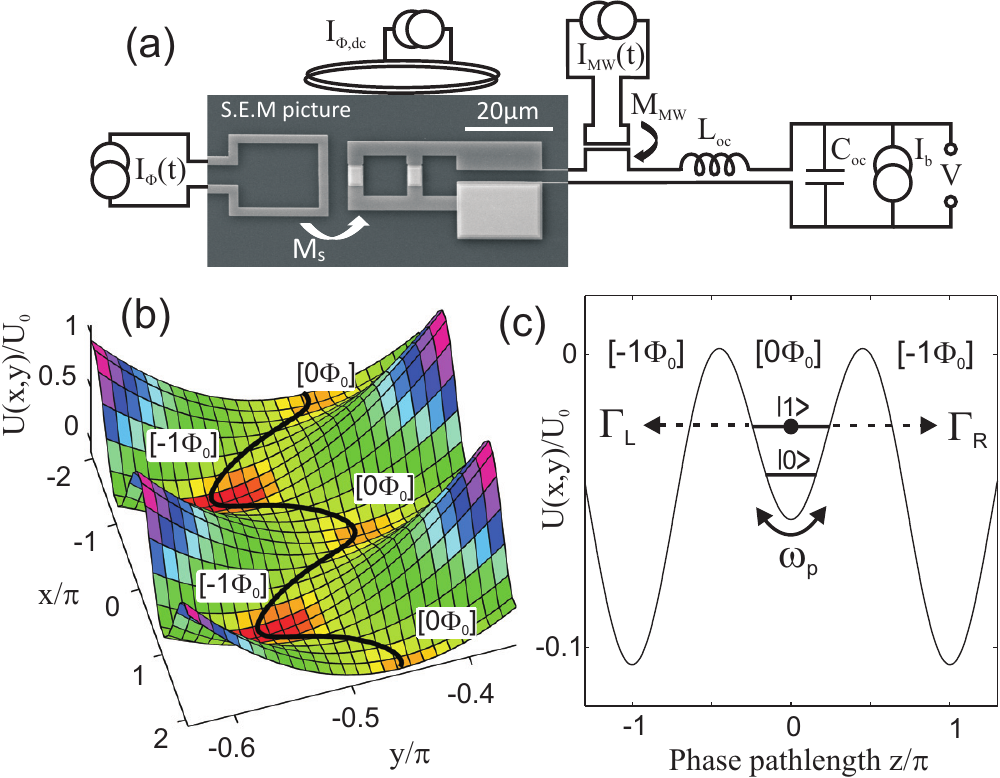}
\caption{\label{fig1} Schematic of experimental setup and camel-back
  potential.(a) Circuit layout. (b) Full 2-D potential for $b = 3.05$,
  $\eta = 0.72$, $\alpha = 0$, $\Phi_\mathrm{ext} = -0.508\Phi_0, I_b
  = 0$, showing the families of minima associated with the $[0\Phi_0]$
  and $[-1\Phi_0]$ fluxoid states. The black line follows the minimum
  energy path. (c) Potential along the minimum energy path,
  parameterized by the path length.}
\end{figure}

Superconducting devices, based on the nonlinearity of the Josephson
junction (JJ), exhibit a wide variety quantum phenomena. During the
last decade, inspired by Macroscopic Quantum Tunnelling (MQT) studies
\cite{Leggett_92}, quantum dynamics of the current biased JJ, dc SQUID
and the rf SQUID phase qubit have been extensively studied
\cite{Martinis_PRL02,Claudon_PRL04,
  Cooper_PRL04,Lisenfeld_PRL07,Dutta_AXv08}. In each of these devices,
the dynamics can be described as those of a quantum particle in a
quadratic-cubic potential. The flux qubit system
\cite{Chiorescu_Science_03}, realized by three or four JJs in a loop,
is described by a double well potential. Here we propose to study a
new potential shape called hereafter a ``camel-back'' double barrier
potential, shown in Fig.~\ref{fig1}c.  This potential is obtained using
the dc SQUID circuit shown in Fig.~\ref{fig1}a in a new way.  The
characteristics of the camel-back potential, including depth and
relative barrier height, are controlled by the SQUID current bias
$I_b$ and flux bias $\Phi_\mathrm{ext}$. There is a special line we
call the ``optimal line'' in these two bias parameters at which the
barrier heights are equal and anharmonicity is quartic. Because of the
symmetry of the potential, the system can escape from the central well
via tunneling through either of the two barriers to an adjacent deeper
well. We investigate this double path escape and analyse it using MQT
theory.  When the depth of the central well is adjusted so that the
escape rate is negligible but anharmonicity is significant, the two
lowest energy levels $\left|0\right>$ and $\left|1\right>$ constitute
a phase qubit. We show that this qubit is insensitive to dephasing due
to current bias fluctuations on the optimal line.

%We control the depth of the central well of the camel-back potential
%on a nanosecond time-scale via $\Phi_\mathrm{ext}$.

A dc SQUID circuit has two degrees of freedom corresponding to the
phase differences $\phi_1$ and $\phi_2$ across its two JJs. The
dynamics are analogous to those of a particle of mass $m =
2C(\Phi_0/2\pi)^2$ in the 2-D potential
\cite{Tesche_JLTP77,Seguin_PRB92}
\begin{equation}
  \label{eq:1}
  \begin{array}{llc}
%  \frac{U(x,y)}{U_0} & = & -\cos{x}\cos{y}
%  - sx + b\left(y-y_b\right)^2
%  \\ & & -\alpha\sin{x}\sin{y}-\eta{}sy.
  U(x,y) & = & U_0[-\cos{x}\cos{y}
  - sx + b\left(y-y_b\right)^2
  \\ & & -\alpha\sin{x}\sin{y}-\eta{}sy].
  \end{array}
\end{equation}
Here $x = (\phi_1+\phi_2)/2$ and $y = (\phi_1-\phi_2)/2$.  Fixed for a
given sample are the Josephson energy
$U_0=(I_{c1}+I_{c2})\Phi_0/2\pi$, the junction to loop inductance
ratio $b = \Phi_0/2\pi LI_c$, the critical current asymmetry $\alpha =
(I_{c2}-I_{c1})/2I_c$, and the loop inductance asymmetry $\eta =
(L_2-L_1)/L$. Here $I_c = I_{c1}+I_{c2}$, $I_{c1}$ and $I_{c2}$ are
the critical currents of the two junctions, $L_1$ and $L_2$ are the
geometric inductances of the two arms of the SQUID loop, $L =
L_1+L_2$, $C$ is the capacitance of each junction, and $\Phi_0 = h/2e$
is the quantum of flux. The external control parameters $I_b$ and
$\Phi_\mathrm{ext}$ enter into the potential through
$y_b=\pi\Phi_\mathrm{ext}/\Phi_0$ and $s = I_b/I_c$. For our sample,
$I_c = 11.22~\mathrm{\mu{A}}$, $C = 250.3~\mathrm{fF}$, $b = 3.05$,
$\eta = 0.72$, and $\alpha = 0.0072$.

The first term in $U(x,y)$, due to the junctions, describes a 2-D
periodic array of minima and maxima. This array can be tilted in the
$x$-direction with an applied current bias. Magnetic field energy
associated with circulating current gives rise to the parabolic term
in the $y$-direction, the minimum of which is shifted by the external
flux.

Stable, stationary states of the system correspond to minima of
$U(x,y)$. There can exist one, two, or more minima families
corresponding to distinct fluxoid states $[n\Phi_0]$. For each, when
$s$ exceeds a flux dependent critical value $s_c[n\Phi_0](y_b)$, the
related local minima disappear. For small values of $b$, the parabolic
term in $U(x,y)$ is shallow, and there can be many fluxoid states. For
$b \gg 1/\pi$, as in our case, the parabolic term is steep and there
is only one stable fluxoid state except in a small region around
$\Phi_\mathrm{ext}/\Phi_0 = 0.5 \pmod{1}$, $I_b \approx 0$ where there
are two states with opposite circulating current. Hereafter we will be
focusing on this region.

In general, dynamics is described by 2-D motion in the potential. In
our case, the particle moves through a valley in which the curvature
is much larger in one direction ($\sim$100 GHz) than the other (10-20
GHz). To a good approximation therefore, motion is one dimensional
along the path of minimum curvature which connects minima and saddle
points. We parametrize this path with the phase length~$z$ (see black
line in Fig.~\ref{fig1}b). $U(z)$ in Fig.~\ref{fig1}c depicts the
``camel-back'' potential shape we are investigating. In a typical
experiment, the system is initialized in the central well, which
corresponds to the [0$\Phi_0$] fluxoid state. The deeper wells on
either side of the central well both correspond to the [-1$\Phi_0$]
fluxoid state. Starting from the central well, the system can escape
via tunneling through the barriers in either of the two physically
distinct directions to the [-1$\Phi_0$] fluxoid state.

In the perfectly symmetric case, the potential near the central
minimum will be harmonic with a quartic perturbation. More generally,
the Hamiltonian for small oscillations in $U(z)$ is
\begin{equation}
  \label{eq:2}
    H = \frac{1}{2}\hbar\omega_p(\hat{P}^2+\hat{Z}^2)
  -\sigma\hbar\omega_p\hat{Z}^3
  -\delta\hbar\omega_p\hat{Z}^4.
\end{equation}
Here $\omega_p$ is the zero amplitude oscillation frequency in the
direction of minimum curvature, and $\hat{Z}=z \sqrt{m\omega_p/\hbar}$
and $\hat{P}=p/\sqrt{\hbar\omega_p m}$ are the reduced position and
corresponding momentum operators. Treating the anharmonic terms as
perturbations, to second order the transition energy between levels
$n-1$ and $n$ is $\hbar\omega_{n-1,n} = \hbar\omega(1-n\Lambda)$,
where the anharmonicity is $ \Lambda = \frac{15}{2}\sigma^2+3\delta$
\cite{Landau-Lifschitz_1991}.  We have calculated the escape
probability for the camel-back potential with a double escape path in
the quantum limit using the instanton formalism~\cite{Callan_PRD77}.
For a duration $\Delta{t}$, it reads
$P_\mathrm{esc}(I_b,\Phi_\mathrm{ext}) = 1 -
\mathrm{e}^{-(\Gamma_R+\Gamma_L)\Delta{t}}$, where $\Gamma_{R,L} =
A_{R,L}\omega\sqrt{N_{R,L}}\,\exp\!\left[-B_{R,L} N_{R,L}\right]$.
Here $R$ and $L$ refer to the right and left barriers.  $N_{R,L} =
\Delta{U_{R,L}}/\hbar\omega$ are the normalized barrier heights.  The
general expression of the coefficients $A_{R,L}$, and $B_{R,L}$
depends on the potential shape.  In the symmetric case where
$\sigma(I_b,\Phi_\mathrm{ext})=0$, the potential is quadratic-quartic,
$A_{R,L} = 2^{\frac{5}{2}}\pi^{-\frac{1}{2}}$ and $B_{R,L} = 16/3$.
Far from this symmetric line the potential is quadratic-cubic, the
escape rate through one barrier is dominant (\textit{e.g.}
$\Gamma_L=0$), and we retrieve the standard MQT situation
($\delta=0$): $A_R = 6^{\frac{3}{2}}\pi^{-\frac{1}{2}}$ and $B_R =
36/5$~\cite{Leggett_92}.

\begin{figure}[htb*]
\includegraphics[width=3.5in]{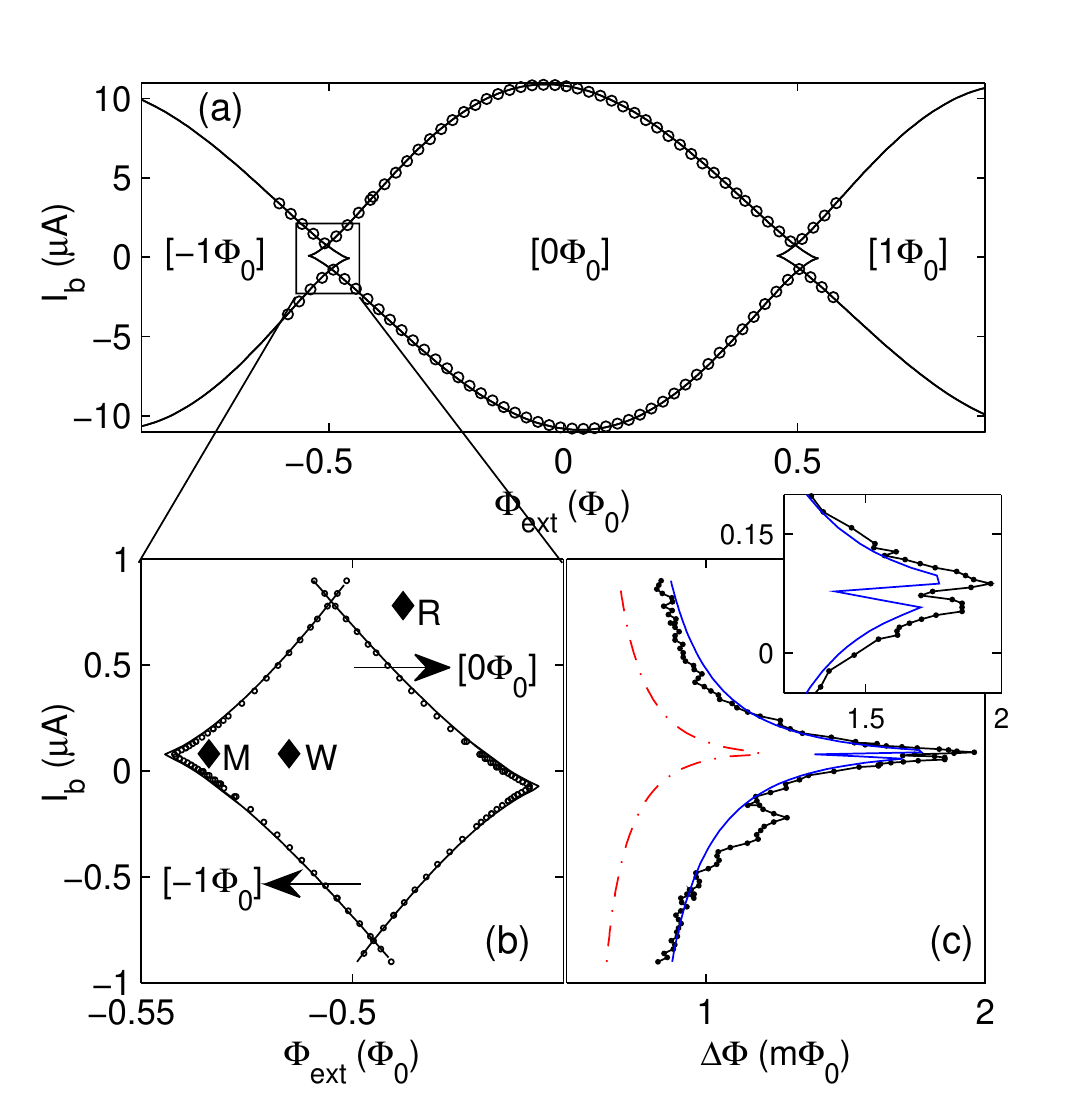}
\caption{\label{fig2} Escape from the ground state. (a) Critical lines
  of three neighboring fluxoid states, measured (symbols), and
  standard MQT theory fit (lines), denoting $I_{50\%}$, the amplitude
  of a 60 $\mu$s $I_b$ pulse that yields $P_\mathrm{esc}=50\%$ to
  the voltage state of the SQUID. (b) Critical lines representing 50\%
  escape out of fluxoid states [0$\Phi_0$] and [-1$\Phi_0$] in the
  region around $\Phi_\mathrm{ext}=-\Phi_0 /2$ due to a 100 ns flux
  pulse, measured (symbols), and the generalized MQT theory fit
  (lines). The [0$\Phi_0$] and [-1$\Phi_0$] fluxoid states are both
  stable in the central region enclosed by the critical lines. Escape
  is either to the SQUID voltage state ($I_b > 0.8~\mu$A) or to the
  adjacent fluxoid state ($I_b < 0.8~\mu$A). The points W, M, and R
  indicate the Working point, quantum Measurement point and Readout
  point for a typical camel-back potential phase qubit experiment. (c)
  Width of ground-state escape $\Delta\Phi$: measurements
  (points+lines), and generalized MQT theory with (solid line) and
  without (dashed line) 9 nA RMS low-frequency current noise. The
  location of the dip near the maximum $\Delta\Phi$ corresponds to
  the point where symmetry leads to a reduction in sensitivity to
  noise.}
\end{figure}

A schematic of our experimental setup is shown in
Fig.~\ref{fig1}a. Our sample was fabricated at PTB using a
Nb/AlO$_x$/Nb trilayer process with SiO$_2$ dielectric and a critical
current density of 300 A/cm$^2$~\cite{Dolata_JAP2005}. The 5
$\mu{m}^2$ junctions are embedded in a square loop with inner size
10x10 $\mu{m}^2$. An off-chip coil provides a dc flux bias. Current
bias and voltage leads, heavily filtered at various stages of the
cryostat\cite{Fay_thesis}, connect at the right of the SQUID.  Fast
flux pulses are inductively coupled via the on-chip loop to the left
of the SQUID. Microwave (MW) excitation is applied via an on-chip loop
which couples inductively to the current bias leads. The MW excitation
must be in {\it current}, rather than flux, because for the symmetric
camel potential, small amplitude oscillations occur for the most part
in the $x$ direction, and therefore must be excited via the $-sx$ term
in $U(x,y)$. The fast flux and MW excitation lines are $50~\Omega$
coaxial with -20 dB attenuators at 1 K and base temperature. The SQUID
chip is enclosed in a small copper box thermally anchored to the
mixing chamber of a dilution refrigerator with a base temperature of
30 mK. The cryostat is surrounded by a superconducting Pb shield,
inside a $\mu$-metal shield, inside a soft iron shield.

Fig.~\ref{fig2}a shows the switching current $I_{50\%} \simeq
I_cs_c[n\Phi_0](y_b)$ as a function of flux for the [-1$\Phi_0$],
[0$\Phi_0$], and [1$\Phi_0$] fluxoid states. The interior of each
curve is the region where the corresponding flux state is stable. The
measurements shown in Fig.~\ref{fig2}a were obtained with a standard
technique in which $I_b$ pulses of varying amplitude are applied and a
dc voltage detected across the SQUID when it switches to its voltage
state. With this scheme, however, there is no direct indication of
multiply stable flux states. In Fig.~\ref{fig2}b we use a novel
technique to measure the overlapping critical lines of [0$\Phi_0$] and
[-1$\Phi_0$] flux states close to
$\Phi_\mathrm{ext}/\Phi_0=-0.5$. These two interior critical lines
represent transitions between the two flux states, rather than
transitions to the voltage state, which is why the standard technique
does not detect them.

Our novel escape measurement method proceeds as follows. First, if
necessary, the system is initialized in the desired flux state with an
adiabatic pulse on the fast flux line. $I_b$ is brought to its working
point value.  A flux pulse $\delta\Phi$ is applied via the fast line
for a fixed nanosecond-scale duration, bringing the total externally
applied flux to a ``measurement point'' close to the critical
line. This has the effect of reducing the heights $\Delta{U}_{R,L}$ of
the two potential barriers. $P_\mathrm{esc}$ via tunneling from the
central well through the barriers to the neighboring deeper wells is
thereby greatly increased. The system is brought back to a flux at
which both fluxoid states are stable. The fluxoid state is then read
out via a slow $(\sim{}10~\mu{s})$ $I_b$ pulse. This $I_b$ pulse
brings the system outside the critical line of fluxoid state
[-1$\Phi_0$] but well within that of [0$\Phi_0$]. If the system is in
state [-1$\Phi_0$], it will switch, producing a voltage which is
detected. If it is in state [0$\Phi_0$], it will not switch. We
achieve a one-shot discrimination between flux states of 100\% with
this readout.  The process is completed by bringing $I_b$ to zero and
waiting ~100 $\mu$s for the heat generated by a switching event to
dissipate before repeating. Multiple repetitions, at a rate of about 5
kHz, yield $P_{\mathrm{esc}}$.

The overlapping $P_\mathrm{esc} = 50\%$ critical lines seperating the
$[-1\Phi_0]$ and $[0\Phi_0]$ fluxoid states are plotted in
Fig.~\ref{fig2}b. Each ends in a cusp at the extreme value of flux
where the corresponding fluxoid state is stable. These cusps occur at
a non-zero current bias $I_b^\mathrm{cusp} = \pm\alpha I_c = \pm 81$
nA due to the critical current asymmetry $\alpha$. The horizontal
separation of the cusps scales precisely with $1/b$. Our generalized
MQT theory is accurately able to reproduce the measured data of fig
\ref{fig2}b. Of the parameters that go into this theory, $b$ and
$\alpha$ are treated as free parameters in this fit, $I_c$ and $\eta$
are determined by the fit in Fig.~\ref{fig2}a and $C$ is determined by
a fit to spectroscopic data.

Along the critical line of a given fluxoid state, for $I_b$ above or
below the value $I_b^\mathrm{cusp}$, the potential is tilted to the
right or to the left, and escape occurs preferentially in that
direction. At $I_b^\mathrm{cusp}$, the camel potential is symmetric around the
minima ($\sigma=0$), the two potential barrier heights are equal, and
escape occurs with equal probability in either direction. The cusps in
Fig.~\ref{fig2}b correspond therefore to a double-path
escape.

The width of the escape process contains additional information about
the dependence of the potential on the bias parameters, and on
fluctuations in the bias parameters \cite{Claudon_PRB07}. In
Fig.~\ref{fig2}c, we plot the width $\Delta\Phi =
|\Phi_{80\%}-\Phi_{20\%}|$, as a function of $I_b$. This plot peaks
around $I_b^\mathrm{cusp}$, except that at this point there is a sharp
dip (see insert).  This behavior is explained by double-path MQT
escape if we include low frequency current fluctuations. In this
circuit thermal fluctuations are expected in $I_b$, which we estimate
to be on the order of 10 nA RMS by the equipartition theorem
$\frac{1}{2}kT = \frac{1}{2}LI_\mathrm{RMS}^2$, where $k$ is
Boltzmann's constant, $T \simeq 40$ mK is the circuit temperature, and
$L \simeq 10$ nH is the series isolating inductance. Because of this
noise, the escape probability is averaged:
$\left<P_\mathrm{esc}(I_b,\Phi_\mathrm{ext}) \right>$. The angle
brackets represent a convolution with the probability distribution of
$I_b$, which we assume to be Gaussian with standard deviation
$I_\mathrm{RMS}$. As shown in Fig.~\ref{fig2}c, the addition of
$I_\mathrm{RMS} = 9$ nA is accurately able to explain both the
increase in the overall width, and the presence of a distinctive dip
at $I_b^{cusp}$ which is a result of symmetry in escape direction. The
presence of the dip and our ability to reproduce it with MQT theory is
a striking confirmation of double path escape and low frequency $I_b$
fluctuations in our sample.

\begin{figure}[htb*]
\includegraphics[width=3.5in]{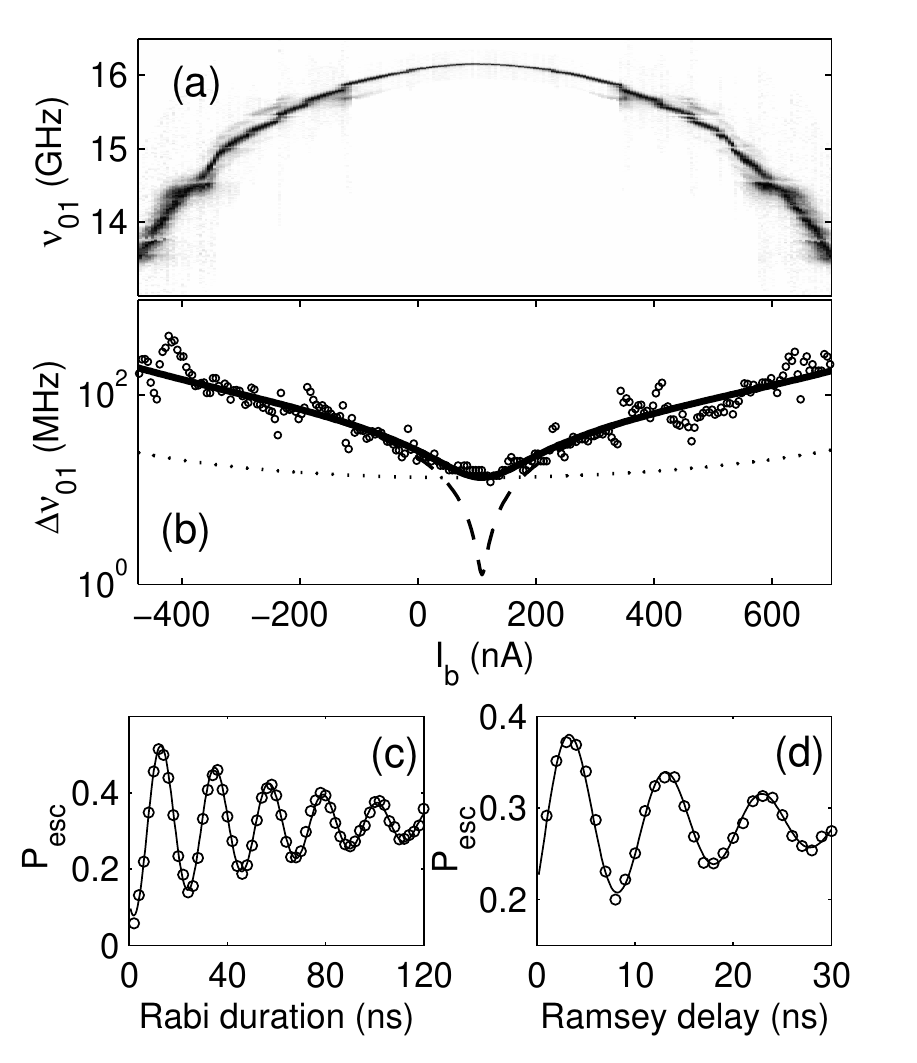}
\caption{\label{fig3} (a) P$_\mathrm{esc}$ versus $I_b$ and MW
  frequency. Dark and bright grayscale correspond to high and small
  P$_\mathrm{esc}$.  (b) Width of the resonance on a semi-log
  scale. The dashed line is the predicted contribution due to 9 nA RMS
  low-frequency current noise. The dotted line is for 40 $\mu\Phi_0$
  RMS low-frequency flux noise. The sum of these two contributions,
  the solid line, accurately reproduces the data (symbols). Rabi (c)
  and Ramsey (d) oscillations at the optimal line at
  $I_b=-71$ nA, $\Phi_\mathrm{ext}/\Phi_0=-0.468$ for the $[-1\Phi_0]$
  flux state.}
\end{figure}

In Fig.~\ref{fig3}a and b we investigate the operation of a camel-back
potential phase qubit corresponding to the two lowest levels
$\left|0\right>$ and $\left|1\right>$ of the anharmonic central well
related to the $[0\Phi_0]$ flux state (see fig \ref{fig1}c). For these
measurements we use the same procedure as for the ground-state escape
measurements except that before the nanosecond measurement pulse, an
adiabatic flux pulse brings the system to the working point flux
$\Phi_W$ where a MW pulse is applied to the fast current line. At the
working point, the barriers are high enough that $P_\mathrm{esc}$ is
negligible.  Immediately following the MW pulse we apply a 5 ns pulse
which projects the qubit state onto the flux state of the SQUID.  This
is possible because $P_\mathrm{esc}$ depends exponentially on the
excitation level of the qubit. The amplitude of the measuring flux
pulse is tuned such that escape will occur with high probability if
the qubit is excited, and low probability if it is not. The
measurement pulse transfers the quantum states $\left|0\right>$ and
$\left|1\right>$ of the qubit to the classical fluxoid states
$[0\Phi_0]$ and $[-1\Phi_0]$ of the SQUID.  Readout of the fluxoid
state, which is itself stable, reveals the projected qubit state, and
repetition yields $P_\mathrm{esc}$.

$P_\mathrm{esc}$ was measured as a function of MW frequency $\nu$ and
$I_b$. Because the MW pulse duration, 800 ns, is much longer that the
relaxation time $T_1 \simeq 100$ ns, the system reaches a steady
state. A peak in $P_\mathrm{esc}$ appears when $\nu$ matches the qubit
transition frequency $\nu_{01}$. Fig.~\ref{fig3}a shows $\nu_{01}$ as
function of $I_b$. It reaches a maximum at
$I_b^\mathrm{op}(\Phi_\mathrm{ext})$ which corresponds to the camel
potential symmetric point. Note that this optimal point is a function
of flux, and is terminated by the cusp at the critical line. This data
was taken at $\Phi_\mathrm{ext} = -0.503 \Phi_0$ for the $[0\Phi_0]$
fluxoid state. Apparent in this spectroscopic image are avoided level
crossings with what are likely microscopic two-level fluctuators, as
first observed by Ref.\cite{Cooper_PRL04}. We observe on average 20
crossings per GHz.  In Fig.~\ref{fig3}b, the spectroscopic width of
the $\nu_{01}$ transition $\Delta\nu_{01}$ is plotted as a function of
$I_b$. A sharp minimum is observed at $I_b=108$ nA, corresponding to
the flat maximum in $\nu_{01}$.

We find that we can accurately model $\Delta\nu_{01}(I_b)$ with a
combination of low-frequency current and flux fluctuations. Because
$\nu_{01}$ depends on the bias parameters, fluctuations cause
$\nu_{01}$ to vary from repetition to repetition, smearing out the
observed resonance. Assuming a Gaussian fluctuation distribution, the
predicted variance in $\nu_{01}$ is
$\left(\frac{\Delta\nu_{I}}{2}\right)^2 =
\left(\frac{\partial\nu}{\partial{I_b}}\right)^2 I_\mathrm{RMS}^2 +
\frac{1}{2} \left(\frac{\partial^2\nu}{\partial{I_b}^2}\right)^2
I_\mathrm{RMS}^4$, for current fluctuations alone, and
$\left(\frac{\Delta\nu_{\Phi}}{2}\right)^2 =
\left(\frac{\partial\nu}{\partial{\Phi_\mathrm{ext}}}\right)^2
\Phi_\mathrm{RMS}^2$ for flux fluctuations alone. Here $\Delta\nu_I$
has been expanded to second order in $I_\mathrm{RMS}$ since
$\frac{\partial\nu}{\partial{I_b}}$ is zero at the optimal line. In
Fig.~\ref{fig3}b, the predicted $\Delta\nu_I$ is plotted as a dashed
line for $I_\mathrm{RMS}= 9$ nA, precisely the same current
fluctuation amplitude used in Fig.~\ref{fig2}c. The dotted line plots
$\Delta\nu_\Phi$ for $\Phi_\mathrm{RMS} = 40~\mu\Phi_0$. The solid
line is the combined prediction $\Delta\nu =
\sqrt{\Delta\nu_I^2+\Delta\nu_\Phi^2}$.  The dashed line is obscured
behind the solid line except in a small region around the optimal
current.  This plot vividly demonstrates the idea of the optimal line:
the effects of current bias fluctuations, which accurately account for
the spectral width away from the optimal line, are rendered negligible
on the optimal line. The residual spectroscopic width, about 10 MHz,
can be explained by a flux noise of $40~\mu\Phi_0$ RMS. Since the
decoherence time $T_2$ scales inversely with $\Delta\nu_{01}$, this
optimal line is also optimal for qubit operations.  Along this line
Rabi and Ramsey oscillations (Fig.~\ref{fig3}c and d) were measured
giving coherence times of $T_\mathrm{Rabi}=67$ ns and
$T_\mathrm{Ramsey}=18$ ns for this current sample.  The anharmonicity
is large enough and the applied power small enough that excitation
beyond the first excited state is negligible, as we have verified by
the linearity of Rabi frequency versus power. The system is confined
to its lowest two levels and can therefore be considered a qubit.

In conclusion, we have studied the quantum dynamics of a novel
quadratic-quartic ``camel'' potential created in a dc SQUID circuit
with $I_b\simeq 0$, $\Phi_\mathrm{ext}\simeq 0.5\Phi_0$. Ground state
escape exhibits critical line cusps and a dip in the escape width
versus bias-current. We explain these two effects with a generalized
double-path MQT escape theory. Moreover due to the particular
potential symmetry, the quantum dynamics is insensitive in first order
to current fluctuations along an optimal line
$I_b^\mathrm{op}(\Phi_\mathrm{ext})$. Along this line, the dc SQUID
can be used as a phase qubit whose main decoherence sources are
residual flux noise and microscopic two-level fluctuators. Future
optimization and exploitation of the unique properties of this system
will aid in the understanding of decoherence mechanisms in quantum
circuits and has the potential to yield a competitive phase qubit.

This work was supported by two ACI programs, by the EuroSQIP and
INTAS projects.


\begin{thebibliography}{}
\bibitem{Leggett_92} {\sl Quantum Tunneling in Condensed Media},
  Modern Problems in Condensed Matter Sciences, Vol. 34, edited by
  Yu. Kagan and A. J. Leggett (Elsevier Science Publishers, 1992).
\bibitem{Martinis_PRL02} J. M. Martinis, S. Nam, J. Aumentado, and
  C. Urbina, Phys. Rev. Lett. {\bf 89}, 117901 (2002).
\bibitem{Cooper_PRL04} K.B. Cooper \textit{et al.},
  Phys. Rev. Lett. \textbf{93}, 180401 (2004).
\bibitem{Claudon_PRL04} J. Claudon, F. Balestro, F. W. J. Hekking, and
  O. Buisson, Phys. Rev. Lett. \textbf{93}, 187003 (2004).
\bibitem{Lisenfeld_PRL07} J. Lisenfeld, A. Lukashenko, M. Ansmann,
  J. M. Martinis, and A. V. Ustinov, Phys. Rev. Lett {\bf 99}, 170504
  (2007).
\bibitem{Dutta_AXv08} S. K. Dutta \textit{et al.}, arXiv 0806.4711
  (2008).
\bibitem{Chiorescu_Science_03} I. Chiorescu, Y. Nakamura,
  C. J. P. M. Harmans, and J. E. Mooij, Science
  {\bf 299}, 1869 (2003).
\bibitem{Tesche_JLTP77} C. D. Tesche and J. Clarke, J. Low Temp.
  Phys. {\bf 29}, 301 (1977).
\bibitem{Seguin_PRB92} V. Lefevre-Seguin, E. Turlot, C. Urbina, D.
  Esteve, and M. H. Devoret, Phys. Rev. B {\bf 46}, 5507 (1992).
\bibitem{Landau-Lifschitz_1991} L. D. Landau and L. M. Lifshitz, {\sl
    Quantum Mechanics: Non-Relativistic Theory} (Course of Theoretical
  Physics, Volume 3), (3ed., Pergamon, 1991).
\bibitem{Callan_PRD77} C. G. Callan and S. Coleman, Phys. Rev. D {\bf
    16}, 1762 (1977).
\bibitem{Dolata_JAP2005} R. Dolata, H. Scherer, A. B. Zorin, and
  J. Niemeyer, J. Appl. Phys. {\bf 97}, 054501 (2005).
\bibitem{Fay_thesis} A. Fay, PhD thesis, Universit\'e Joseph Fourier,
  2008.
\bibitem{Claudon_PRB07} J. Claudon, A. Fay, E. Hoskinson, and
  O. Buisson, Phys. Rev. B \textbf{76}, 024508 (2007).

\end{thebibliography}
\end{document}